\documentclass[11pt]{article}
\usepackage{UF_FRED_paper_style}

\usepackage[utf8]{inputenc}
\usepackage[T1]{fontenc}
\usepackage{graphicx}
\usepackage[export]{adjustbox}
\usepackage{nameref}
\usepackage{hyperref}
\usepackage{enumitem}
\setlist[enumerate]{label*=\arabic*.}
\usepackage{lipsum}  %% Package to create dummy text (comment or erase before start)

\title{Impact of a Blockchain-based Universal Basic Income Pilot: The case of Circles UBI currency
\thanks{Paper prepared for FRIBIS Conference Proceedings 2023}
}

\author{Longo, Alessandro\\% Name author
    \href{mailto:alessandrolongo1@protonmail.com}{\texttt{alessandrolongo1@protonmail.com}} %% Email author 1 
\and Criscione, Teodoro\\% Name author
    \href{mailto:criscione_teodoro@phd.ceu.edu}{\texttt{criscione\_teodoro@phd.ceu.edu}} %% Email author 2
\and Linares, Julio\\% Name author
    \href{mailto:jrlinares16@gmail.com}{\texttt{jrlinares16@gmail.com}}%% Email author 3
\and Avanzo, Sowelu\\% Name author
    \href{mailto:soweluelios.avanzo@unito.it}{\texttt{soweluelios.avanzo@unito.it}}%% Email author 4
    }
    
\date{\today}

\begin{document}
% %%%%%%%%%%%%%%%%%%%%%%%%%%%%%%%%%%%%%%%%%%%%%%%%%%%%%%%%%%
% %%%%%%%%%%%%%%%%%%%%%%%%%%%%%%%%%%%%%%%%%%%%%%%%%%%%%%%%%%
% ABSTRACT
% %%%%%%%%%%%%%%%%%%%%%%%%%%%%%%%%%%%%%%%%%%%%%%%%%%%%%%%%%%
% %%%%%%%%%%%%%%%%%%%%%%%%%%%%%%%%%%%%%%%%%%%%%%%%%%%%%%%%%%
{\setstretch{.8}
\maketitle
% %%%%%%%%%%%%%%%%%%
\begin{abstract}
% CONTENT OF ABS HERE--------------------------------------

% Background
Circles UBI is a blockchain-based Community Currency System (CCS) that has been active in Berlin (Germany) since October 2021. The Circles Coop, which launched the project in 2021, was shut down in December 2023.
% Methods
In this paper, we show the results of a survey carried out between October and November 2023. The questionnaire was articulated in fifty-one questions divided into seven main sections: “Business Information”,  “Economic Network”, “Use of Circles”, “Knowledge and Perception of Circles”, “Circles Events Participation”, “Socio-economic position” and “Satisfaction”. It was distributed via an online form, and when possible, combined with a semi-structured interview - via Zoom or in person. 
% Results
The respondents were twenty-five individuals involved in various ways in the Circles’ network. The main emerging narrative points out how their participation was deeply motivated by their identification with the values and ideals (e.g., "basic income") of the Circles community. Among them, we selected 5 profiles that stood for their difference in type and degree of involvement. Finally, we report some stories of economic linkages that suggest a positive externality in adopting a local community currency.
% Contribution
To our knowledge, this is the first qualitative study of a universal basic income designed as a community currency and adopting blockchain technology. This pilot project was a remarkable experiment for its adopted advanced technological and social innovations. In fact, as far as we know, the integration of "basic income" and "local currency" features has been experimented with only in two other cases (Marica, Brazil and Barcelona, Spain) and none of them adopted a decentralized ledger system. In this work, we try to outline strengths and weaknesses that emerged after about two years of activity. For this reason, future researchers and activists interested in this field will find valuable information.

% END CONTENT ABS------------------------------------------
%% <-- Keywords HERE!
\noindent

\end{abstract}
}

\makeatletter
\newcommand\setcurrentname[1]{\def\@currentlabelname{#1}}
\makeatother

% %%%%%%%%%%%%%%%%%%%%%%%%%%%%%%%%%%%%%%%%%%%%%%%%%%%%%%%%%%
% %%%%%%%%%%%%%%%%%%%%%%%%%%%%%%%%%%%%%%%%%%%%%%%%%%%%%%%%%%
% BODY OF THE DOCUMENT
% %%%%%%%%%%%%%%%%%%%%%%%%%%%%%%%%%%%%%%%%%%%%%%%%%%%%%%%%%%
% %%%%%%%%%%%%%%%%%%%%%%%%%%%%%%%%%%%%%%%%%%%%%%%%%%%%%%%%%%

% --------------------
\section{Introduction}
% --------------------

Circles UBI is a blockchain-based Community Currency System (CCS) launched in Berlin, Germany in 2021, and designed to offer a Universal Basic Income independent of state control. Universal Basic Income is defined by The Basic Income Earth Network (BIEN) as a periodic cash payment delivered to all individuals unconditionally~\citep{miller2020basic}. Circles UBI employs blockchain-based tokenization to provide an unconditional income to its participants in the form of a token named CRC. 

Blockchain technologies’ advancements allow for the creation of entire application systems. In particular, Ethereum facilitates the decentralized execution of Turing-complete smart contracts~\citep{buterin2014next}. This development has significantly broadened the scope of blockchain applications, giving rise to a novel category of software systems known as Decentralized Applications (DApps)~\citep{zhang2018}. 

Circles executes its scope through the Circles DApp, where participants receive their personal currency upon registration, allowing them to spend it as a claim to goods and services within the network. Trust connections between participants are established via a Web of Trust (WoT) protocol, ensuring fungibility among personal currencies~\citep{lepinski2005, buterin2014next}.

The current paper focuses on the first pilot project launched by Circles Coop~\citep{circlescoop} in Berlin between 2021 and 2023, where a subsidy program for local businesses allowed them to convert Circles tokens into national currency (Euro), referred to in the text as “Subsidy Program”. This program tried to address the currency hoarding of businesses that could not spend it in the economic network. In fact, it aims to incentivize businesses to accept Circles tokens as a medium of payment, enhancing the network's overall expansion.

The primary objective of this study is to delve into the impact of the Berlin pilot on the lives of its participants. Our research is the qualitative counterpart to a prior quantitative study of the blockchain dataset underpinning Circles' network~\citep{avanzo2023} using network analysis methodologies. Furthermore, employing a financial-diaries-inspired approach~\citep{rutherford2009}, we conducted interviews with twenty-five participants in the Circles project to garner a comprehensive understanding of their experiences. The detailed survey responses are presented in Section~\nameref{sec:results}, with the complete set of questions provided in the \nameref{sec:appendix}.

Drawing from these findings and given the reduced sample size, we identified five profiles of individuals and businesses that can offer interesting insights into how Circles UBI affected the lives of its participants within the Berlin context. Some of these profiles are enriched with visualization of the transaction data recorded in the blockchain, which validates the value of the insights identified through the interviews. 

Finally, our perspective merges academic interest with a pragmatic and strategic approach: we hope that our findings can help Circles and similar pioneering projects in their future steps.
% --------------------
\section{Methodology}\setcurrentname{Methodology}\label{sec:methods}
% --------------------

\begin{figure}[H]

  \centering
  \includegraphics[scale=0.45, center]{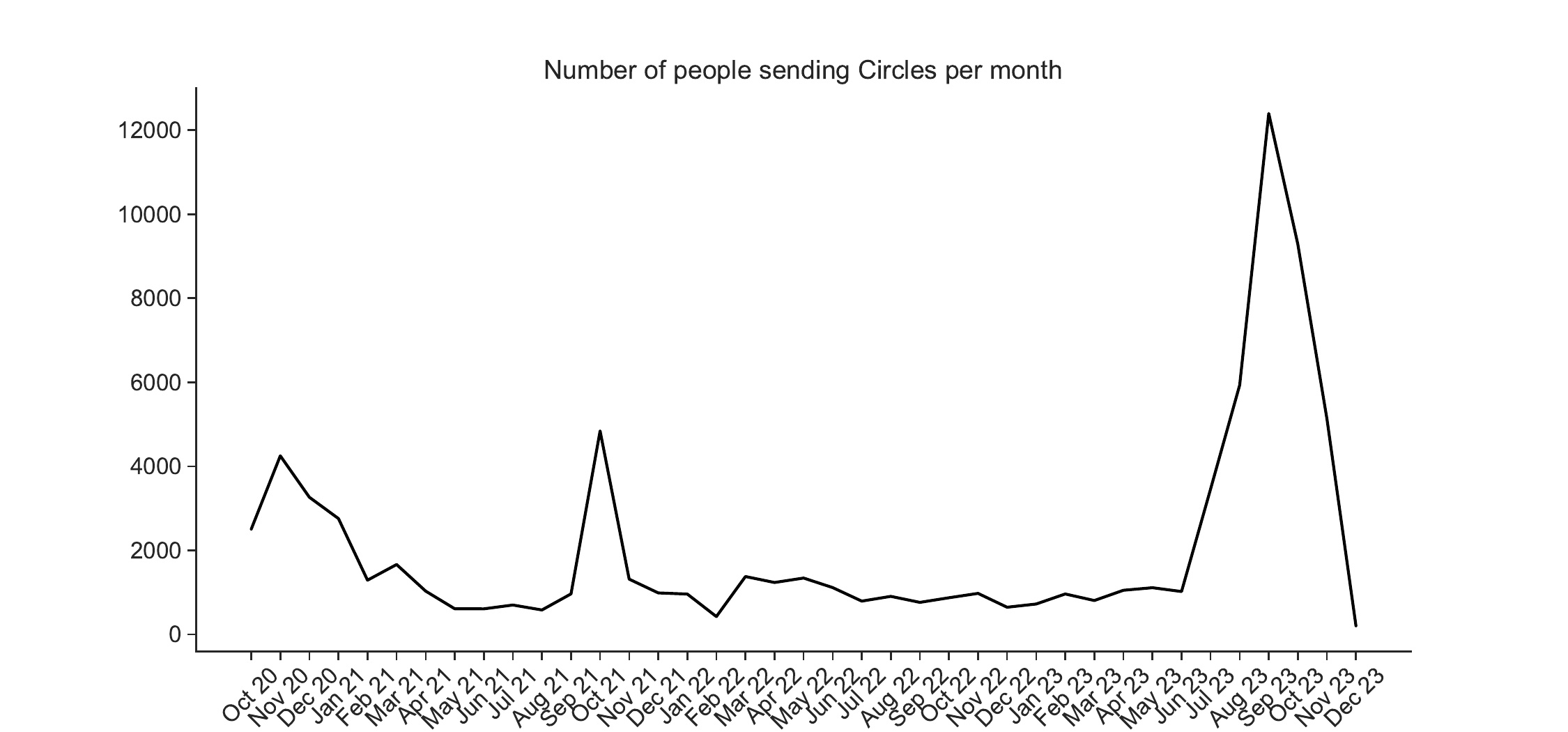}
  \caption{Number of participants sending Circles units per month.}
  
  \label{fig:senders}
\end{figure}

The Circles pilot is an open experiment and it does not adhere to institutional or academic control, characterized by specific participation criteria and limitations on personal expenses, as observed in other studies on Universal Basic Income~\citep{standing2021}. Circles’ network has loose admission and participation criteria and the number of active participants has been stable for most of the entire project’s duration with two peaks in the summer of 2021 and later in 2023 (see Figure \ref{fig:senders}). Hence, the experiment's open nature requires examination within a comparably open research framework. Circles adopts a paradigm at the nexus of innovative technologies and unorthodox monetary theories, making it a singular case of a UBI pilot. Thus, we have opted for a hybrid analysis to assess its impact, blending quantitative and qualitative methods. 

Our approach draws inspiration from existing scholarly work on Universal Basic Income (UBI), adapting the financial diaries approach to align with the unique context of Circles. 
Firstly, Circles' conceptualization of Universal Basic Income departs from the norm. Contrary to conventional discussions that predominantly situate UBI within the purview of state-centric interventions~\citep{torry2019}, Circles posits UBI as a transformative instrument capable of liberating currency from the confines of state control, resonating with innovative experiments in the field, like the dual currency structure proposed by Hornborg~\citep{hornborg2017} and Recivita's pioneering work~\citep{recivitas} in Quatinga Velho, Brazil~\citep{basicincomebrasil}. 

This perspective doesn't position UBI  as a merely economic policy but as a multifaceted tool within a broader arsenal actively challenging prevailing power structures and colonial legacies~\citep{adams2022}. Within Circles project, UBI is framed as a catalyst for establishing a democratic money commons, thereby disrupting traditional power dynamics~\citep{linares2022}. Beyond a mere economic safety net, in this context, UBI represents the autonomy to reject relationships characterized by domination~\citep{widerquist2013}. Circles theoretical frame must be kept in mind when making methodological choices and interpreting the manifold effects of Circles UBI on the lived experiences of its participants. 

Circles UBI shares affinities with a community currency system, similar to existing projects like Real Economy Currency (REC) in Barcelona~\citep{belmonte2021} or Mumbuca in Brazil~\citep{cernev2016}. Yet, none of these projects leverages blockchain and smart contracts to automate and decentralize the system. In this regard, Circles UBI offers a unique case of study.

The financial diaries methodology underpins the qualitative part of our research. Daryl Collins, Stuart Rutherford, and more first developed financial diaries to provide an empirical method for comprehending the financial lives of low-income population segments, presenting the results in Rutherford et al. (2009)~\citep{rutherford2009}. The diaries track the cash flows and decision-making processes of individuals in a household over time. The money flows provide a tool to enhance our understanding of how people living in poverty deal with low and often unpredictable incomes. Financial diaries can complement existing multidimensional metrics as they provide a very granular and locally comprehensive look into specific groups of people, uncovering dynamics that might be hidden in nationwide statistics. 

In our research, we adopted a repurposed version of the financial diaries, drawing inspiration from it for the interviews and the presentation of the results as personal profiles. Through this method, we want to illuminate the effective impact of the UBI offered by Circles on its recipients in Berlin. To achieve this, we contextualize the spending patterns in CRC -i.e., the Circles token - within the broader financial life of respondents, through a survey analysis and semi-structured interviews.

The survey is articulated in fifty-one questions divided into seven main sections: “Business Information” (for business owners),  “Economic Network”, “Use of Circles”, “Knowledge and Perception of Circles” “Circles Events Participation”, “Socio-economic position” and “Satisfaction”. Between October and November 2023, the questionnaire was distributed via an online form. When possible, the questionnaire was combined with a semi-structured interview - via Zoom or in person - where participants were invited to expand their answers to collect more data and impressions on their experiences and perceptions. 

Furthermore, the results are integrated with the analysis of payment data stored in the Circles blockchain system. The recent advancement of payment system digitization has enabled researchers to conduct thorough empirical studies that were previously impossible. The quantitative empirical examinations of digital community currency systems could significantly bolster the resilience of local economies~\citep{criscione2022}. Blockchain's abundance of data, often stored publicly and openly, makes it a valuable resource for research. The transparency and immutability of blockchain records, coupled with decentralized accessibility, offer researchers unique insights across various domains~\citep{bwalya2019}. Such transparency can also be considered an issue for privacy and different actors in the blockchain space are working with privacy-enhancing technologies to tackle it~\citep{circlesentropy}. Since these network analytical tools have been proven to be valid for analyzing highly granular data pools of a local economy, we employed some of those techniques to examine these data~\citep{mattsson2023}. 

% --------------------
\section{Results}\setcurrentname{Results}\label{sec:results}
% --------------------

This section is dedicated to the main presentation of the results of the survey and the challenges that unfolded during the interview process. 

\subsection{Sample}\setcurrentname{Sample}\label{sec:results:sample}

Our sample is composed of twenty-five individuals involved in various ways in the Circles’ network.  The biological sex distribution includes nine female, five male, and eleven non-reporting. The gender distribution within the sample revealed twelve female, eight male, and three gender non-conforming individuals, providing a nuanced understanding of the impact of UBI interventions across diverse gender identities.

The national distribution of the interviewed sample mirrors the cosmopolitan nature of the city of Berlin. Italians constitute the largest national group, with six representatives, followed by Germans with five participants. The sample further encompasses the perspectives of Polish and US nationals, each contributing two interviewees. The inclusion of individuals from Australia, Belarus, Brazil, Britain, Canada, the Netherlands, France, Mexico, Peru, and Spain, with one representative each, highlights the international relevance of UBI interventions and their potential impact on migrant communities. From a demographic point of view, we interviewed seventeen individuals in their thirties, five in their twenties, and one in their forties. The youngest individual is 25 years old and the older is 41 years old. 

Within the Circles’ system, two different kinds of accounts exist: \textit{Individual accounts} and \textit{Shared accounts}. This distinction is reflected in the technical architecture and the monetary policy associated with it. \textit{Individual accounts} are associated with a \textit{Safe Smart Contract}~\citep{safewallet} that holds their personal keys and is designed to be matched with a single, verified identity. These are the accounts that issue the monthly Universal Basic Income once the verification process is completed. \textit{Shared accounts} are a fork of the original \textit{Safe} contracts developed further to be used as collective wallets. They can be created by single participants and other participants can associate with them. These collective accounts don’t issue basic income. In the Berlin pilot, \textit{Shared accounts} have been used by businesses to transact in CRC. 

This is the nature of the \textit{Individual} / \textit{Business} distinction in our research, which is reflected in the questionnaire. Moreover, the sampled businesses in the Berlin pilot were involved in the Subsidy Program, the aforementioned incentive system designed to expand the reach of the pilot. This element further distinguished the two possible kinds of accounts and the different experiences of the Circles network members. Thus, our questionnaire proposed two parallel sets of questions if the person interviewed used an \textit{Individual account}, a \textit{Shared} one, or both. We interviewed fourteen individuals and eleven businesses, of which nine were also active with an \textit{Individual} account. 

\subsection{Spending Patterns}\setcurrentname{Spending Patterns}\label{sec:results:patterns}

\begin{figure}[H]

  \centering
  \includegraphics[scale=0.45, center]{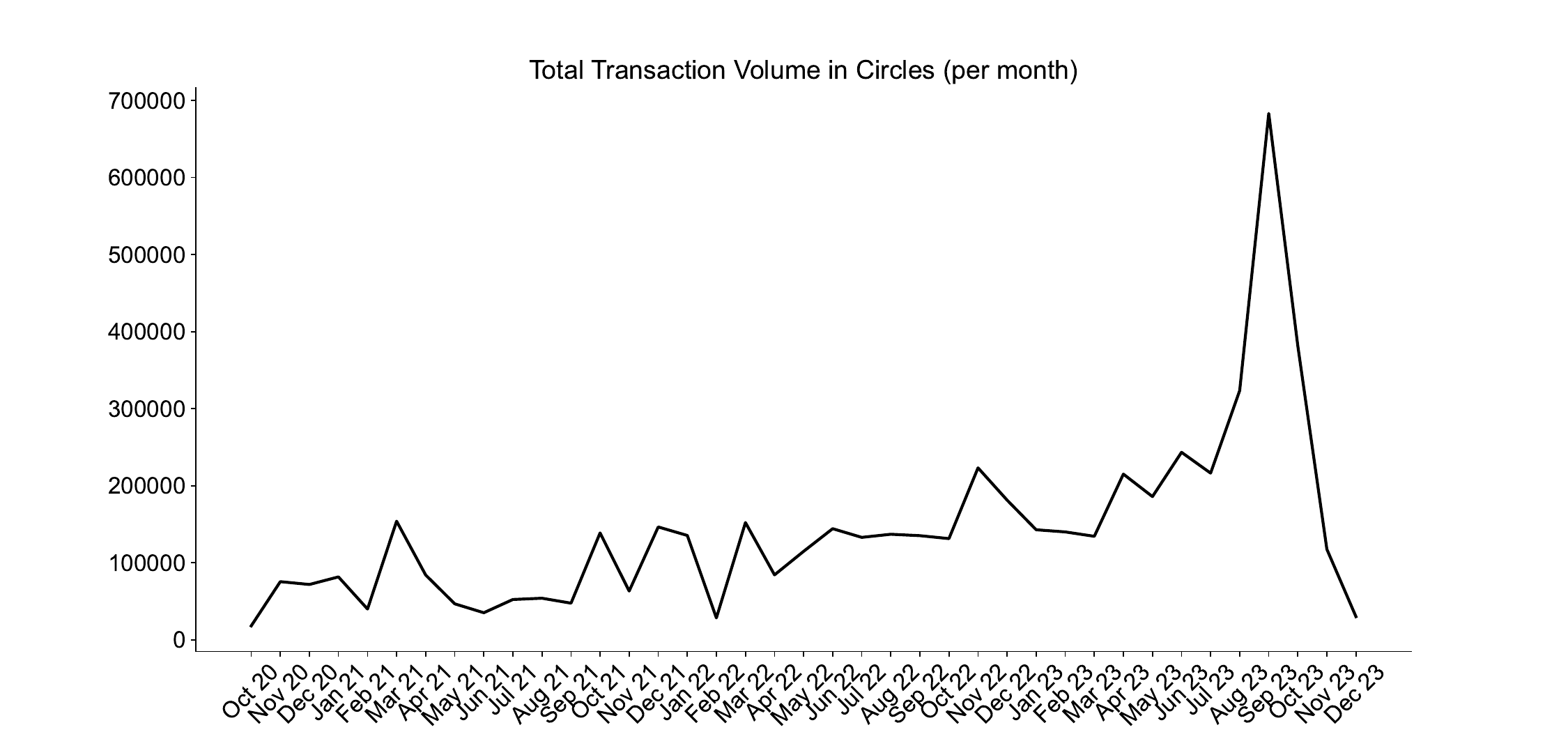}
  \caption{Total transaction volume in Circles units per month.}
  
  \label{fig:transactions}
\end{figure}

An initial insight from the questionnaire concerns the fluctuations in participants’ usage, both for businesses and individuals. Eighteen participants answered affirmatively, explaining how and why their usage changed over time. Seven participants responded negatively as their usage remained constant and two participants did not answer. We discerned different patterns among the seventeen answers that reported changing behaviors. Six individuals reported a relevant increase in their usage of Circles tokens over time. For these participants, the increase in usage is often attributed to the expansion of goods and services available in the Circles marketplace (see Figure \ref{fig:transactions}). From its launch in 2021 until the Coop’s shutdown in late 2023, the Circles network in Berlin witnessed participation and, conversely, abandonment by various affiliated shops. This fluctuation signified variations in the available products. The constant variation of products available in CRC results in an enriched selection for participants, leading to increased token expenditure as the network expands.

Additionally, the reported variations are reflective of a growing familiarity with the network, where participants transitioned from an initially sporadic usage to a more regular and intentional utilization of their Circles tokens. This transition is highlighted by a growing understanding of the intricate nature of the network and the varied composition of this community.
Conversely, an equivalent cohort of 10 respondents share experiences of decreased or variable token usage over time. They point out the challenges listed below.

\begin{itemize}

\item \textit{Technological and communication factors}. Specifically, the slowness and unreliability of the transaction system often discouraged participants. Moreover, in the beginning, the Telegram channel was considered as a not-so-easy communication tool to check products' availability and receive important notifications. 
The first technical issue was a systemic flaw impacting the network's operation, which was slightly improved over time. The second communication issue was slowly solved by integrating the Telegram channel with an Online Marketplace.

\item \textit{Insufficient and/or irregular supply}, such as the periodic unavailability of desired services in the marketplace.

\item \textit{Geographical factors}, namely the physical distance from Berlin, had an impact on the usage of Circles tokens. Given that our sample comprises 20\% Germans and 80\% migrants, it seems plausible to assume that the majority of respondents may have spent longer or shorter periods outside of Berlin than their German peers, and this fact may have had an effect on usage and engagement with the system.

\item\textit{ Subsidy program influence}. Changes in spending habits were also related to the influence of the subsidy program. As mentioned, participation (or the lack of) in the program affects spending patterns.

\item \textit{Demurrage policy} (i.e., negative interest rate). Interview \#25 brings up the changes introduced in June 2022, when the Circles’ token undertook a revaluation policy. The interviewer mentioned this as a factor in decreasing expenditure.

\end{itemize}

\subsection{Consumption Behaviour}\setcurrentname{Consumption Behaviour}\label{sec:results:consumption}

Among different goods, the category of “food and beverages” prominently dominates as the most sought-after one, as indicated by twenty-one out of twenty-five respondents. Health services, handmade products, bike repairs, and clothing were also prominent in participants' preferences.
By offering financial help to its participants, the Circles network may have influenced their purchasing habits. Questions \#24 and \#25 survey this issue. The former asks participants whether Circles allowed them to buy products that they would not have bought otherwise: twenty participants replied positively and only five negatively. The last question asked participants if they recognized a shift in their consumption habits following their Circles experience. Fourteen interviewed Circles participants expressed a perceived change in their consumption habits; while eleven did not experience any change. 

One noticeable initial change is that numerous participants observe Circles expanding the definition of what falls into the Marxian \textit{“realm of necessity”}~\citep{sayers2011} enabling spending on items like organic products, health services (such as acupuncture and massages), pillows, and ceramics that might be deemed non-essential or too expensive otherwise. In this regard, we notice also how Circles fosters an idea of money that moves beyond the paradigm of scarcity~\citep{angehr2019}. 

Secondly, this shift in consumption consists of a greater connection between participants and local businesses, fostering a sense of community and ethical consumption. For instance, participants point out that Circles allows them to afford items they might not purchase otherwise with Euros. Answer \#24 points out how the experience in Circles influenced how to spend EUR for groceries and food. Generally, both interviewed who felt a change and those who did not (Interview \#23) noticed how being part of such an alternative economic network gave them new perspectives on issues such as “market, value, exchange, community in business” and the importance of having values-aligned economic interactions. Participants also expressed gratitude for the affordability of specific items through Circles, the ability to treat themselves more, and a conscious effort to strike a balance between gains and expenditures. Furthermore, the described change in habits can help in incentivizing the consumption of products with lower carbon emissions, as described in previous research~\citep{liberassini}.

The interviews confirm the different meanings that money can assume when designed differently~\citep{linares2022}. Within capitalism, money is entrapped in a specific configuration, centered on the reproduction of capital through the means of commodities, according to the famous Marx’s formula M-C-M~\citep{marx1885}, and the interests’ system which fuels economic growth and resource extraction. The price to pay is the soaring inequality that affects capitalistic societies~\citep{piketty2013} and the ecological destruction of the planet~\citep{hickel2019}. According to its founders, the Circles project aimed to reverse such process and to plan the becoming-common of money~\citep{greco2009, dodd2014, slater2016}, which can be simply described as a plural series of systems “that put money at the service of their own existence, and not the other way around”~\citep{deangelis2017}. In terms of its scope and following the interviews' results, the Circles pilot project stood out as a commendable initial effort of local re-configuration of the monetary relationships.

A different and somehow discrepant aspect related to the design of Circles surfaced in response to both queries, specifically addressing the incentive system devised to encourage participants to maximize their CRC spending. By incentive systems, interviewees refer to the reevaluation of Circles and the introduction of a policy of \textit{demurrage} (i.e., negative interest rate) that the Circles token has undertaken since the end of May 2022~\citep{circlesubi2022}. Until that point, the Circles system issued daily 8 CRC per participant, an issuance rate based on a general poverty line as defined by the Circles Coop \cite{hickel2021}. After the reform, the daily rate became 24 CRC per day (representing each hour of the day), for a total of 720 CRC per month. Additionally, a program of \textit{demurrage} was introduced in the system. “\textit{Demurrage}” is a monetary policy for the built-in pre‐programmed depreciation of the nominal value of a currency, similar to negative interest, and it has been studied in the context of Complementary Currencies~\citep{godschalk2012}. The Cooperative decided to issue a 7\% rate of demurrage that “acts like a parking fee on money, set against the constant issuance of UBI”~\citep{circlesubi2022}. This change of policy aimed to tackle possible hoarding of tokens and to direct a gradual redistribution of all CRC in the system over time, to avoid pyramid-like schema accumulation, common in cryptocurrency systems~\citep{mukherjee2021}. 

Interview \#15 considered this mechanism as an interesting alternative to the necessity of accumulation typical of market economies. On the other hand, Interview \#12 observed a subtle pressure to buy things they might not necessarily need, describing it as a way to make Circles more "convenient" for their shop. This observation aligns with occasional instances of over-consumption noticed in participants' behavior; moreover, the same interview also highlighted how the perceived precarity of the Berlin pilot throughout 2023 led to faster expenses. This issue contrasts studies relating UBI programs to lower patterns of consumption~\citep{coote2021} but it has to be contextualized within Circles’ offering, which is focused on local and sustainable businesses, thus balancing possible dynamics of over-consumption. 

\subsection{Knowledge and Perception}\setcurrentname{Knowledge and Perception}\label{sec:results:knowledge}

This segment of the questionnaire investigated people’s knowledge and perception of the Circles initiative and what level of awareness and involvement emerged among its participants. The first question asked how individuals got to know Circles: sixteen indicated friends as the main source, five indicated colleagues, three social media, and one specified the MOOS community in Berlin~\citep{moos}. Generally, personal contacts were more functional than social media promotion. 

The following pair of questions asked participants what aspects of Circles ignited their interest at first and in what they were more interested at the time of the interviews. The questionnaire presented three main options, namely the community, UBI, or Blockchain integration. Participants could choose more than one and/or indicate their preferences. The first question’s results are the following: seventeen participants mentioned the UBI as the main motivation to join  Circles, fifteen mentioned the community, and only seven mentioned blockchain. Moreover, a minority of interviewees pointed out how Circles’s values of equality and post-capitalism resonated with their sparking interest. After they participated in Circles, the community was mentioned twenty times as the most interesting aspect, succeeded by UBI with nineteen mentions and blockchain systems with only five. Other factors mentioned are again the anti-capitalistic values and the implementation of the Web of Trust protocol as an anti-sybil solution. 

The next question asked how participants perceived Circles. The results are the following: the project was mainly perceived as a Universal Basic Income with eighteen mentions, followed by fifteen mentions for “means of payment”, fourteen for “means of exchange”, eleven for a “political tool”, five for a “marketing tool”, two for an “investment / financial asset”, one for a “store of value/saving tool” and zero for a “fake coin”.

We can notice the peculiar position of Circles. Despite being powered by blockchain technologies, Circles stands out as a unique case where the innovative technological infrastructure did not play a prominent role in attracting the participants’ interests. According to our interviews, interest in blockchain diminished over the two years of the project, and one participant expressed uncertainty about the utility of blockchain in Circles. This outcome underscores that the innovative potential of Circles lies in its community and the aspiration to forge an alternative economic circuit grounded in shared values. This observation also highlights a fundamental aspect of blockchain technologies. Rather than being viewed solely as trustless systems, they should be regarded as confidence machines~\citep{defilippi2020}. More specifically, the reliability of blockchain computing mechanisms is contingent on the presence of effective governance structures and a vigilant community that supports and upholds their functionality. This aligns with Robert Putnam's concept of social capital~\citep{haeuberer2011}, where the strength and resilience of a community depend on the bonds, trust, and shared values among its members. The success of projects like Circles depends not only on the technology they employ but also on the strength of the social bonds and networks they create. 

Diverse perceptions translated into different levels of involvement among participants. In response to the question about participating in Circles events, twenty-one individuals confirmed their involvement, while four indicated a negative response. Delving deeper into event attendance preferences, fourteen participants typically attended both the Market and the Assembly, seven exclusively engaged in the Market, and four chose not to disclose their preference.

Further exploring the frequency of their event attendance, eight interviewees reported participating once per quarter, another eight attended once per month, five engaged approximately once per year, and four respondents opted not to respond. Then, we asked what kind of activities participants got involved in during Circles’ events. The main activity mentioned is the “purchase of goods” with seventeen mentions, “looking for new friends” holds fourteen mentions, “participation at the assembly” is mentioned twelve times, “selling goods” is mentioned eleven times (as we interviewed eleven businesses) and finally, with ten mentions, we find the “increase your trust network / looking for new potential business partners” option.

We must notice how our sample was composed of what may be defined as one of the “core groups” of the Pilot program in Berlin - composed mostly of many associated businesses and long-time participants. Consequently, the nature of their participation is biased by this role.  However, it is arguable that the community aspect emerged as a central driver, emphasizing the importance of shared values in constructing a parallel economic circuit. Money is the product of a set of social relationships and to achieve a different reconfiguration of it, the role of the totality of social relationships (the community) is very important.

However, the degree of participation and involvement was also fluctuating, showing how Circles allowed for different degrees of participation, as several participants only joined passively the community, adopting the role of “aleUBI recipients”. This position within the Circles system will be analyzed in Section~\nameref{sec:profiles:casual}.

Some of the qualitative observations are corroborated by network analysis. The Jaccard similarity index (JSI) indicates how much similar two sets are. It is calculated by the ratio between their intersection set over their union set. We considered the similarity between senders and receivers of Circles for each interviewee and their trusted connections.

\begin{figure}[H]

  \centering
  \includegraphics[scale=0.45, center]{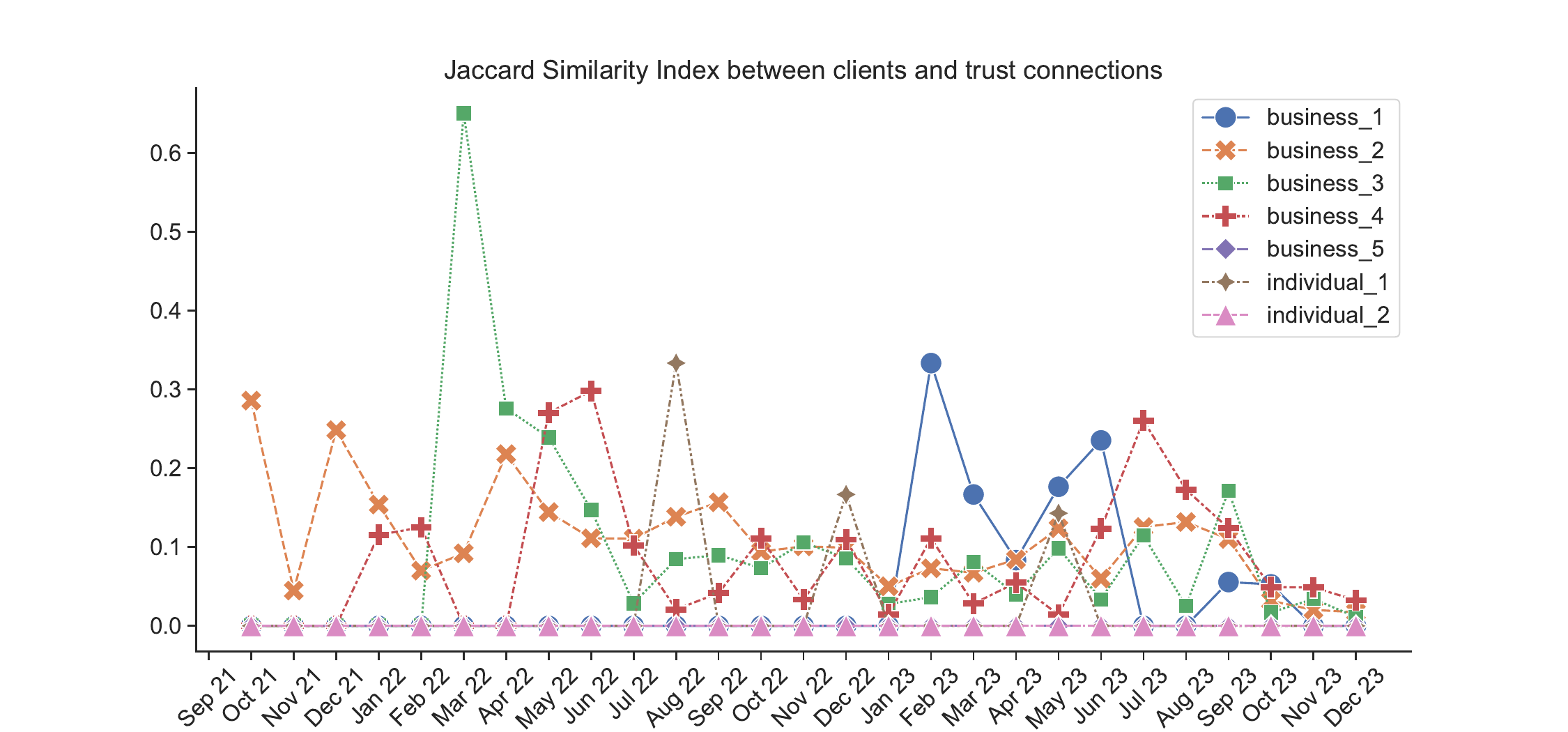}
  \caption{Jaccard Similarity Index between clients and trust connections of selected profiles. In the case of non-business individuals, \textit{clients} are those from whom such individuals receive Circles.}
  
  \label{fig:jaccard_clients}
\end{figure}

\begin{figure}[H]

  \centering
  \includegraphics[scale=0.45, center]{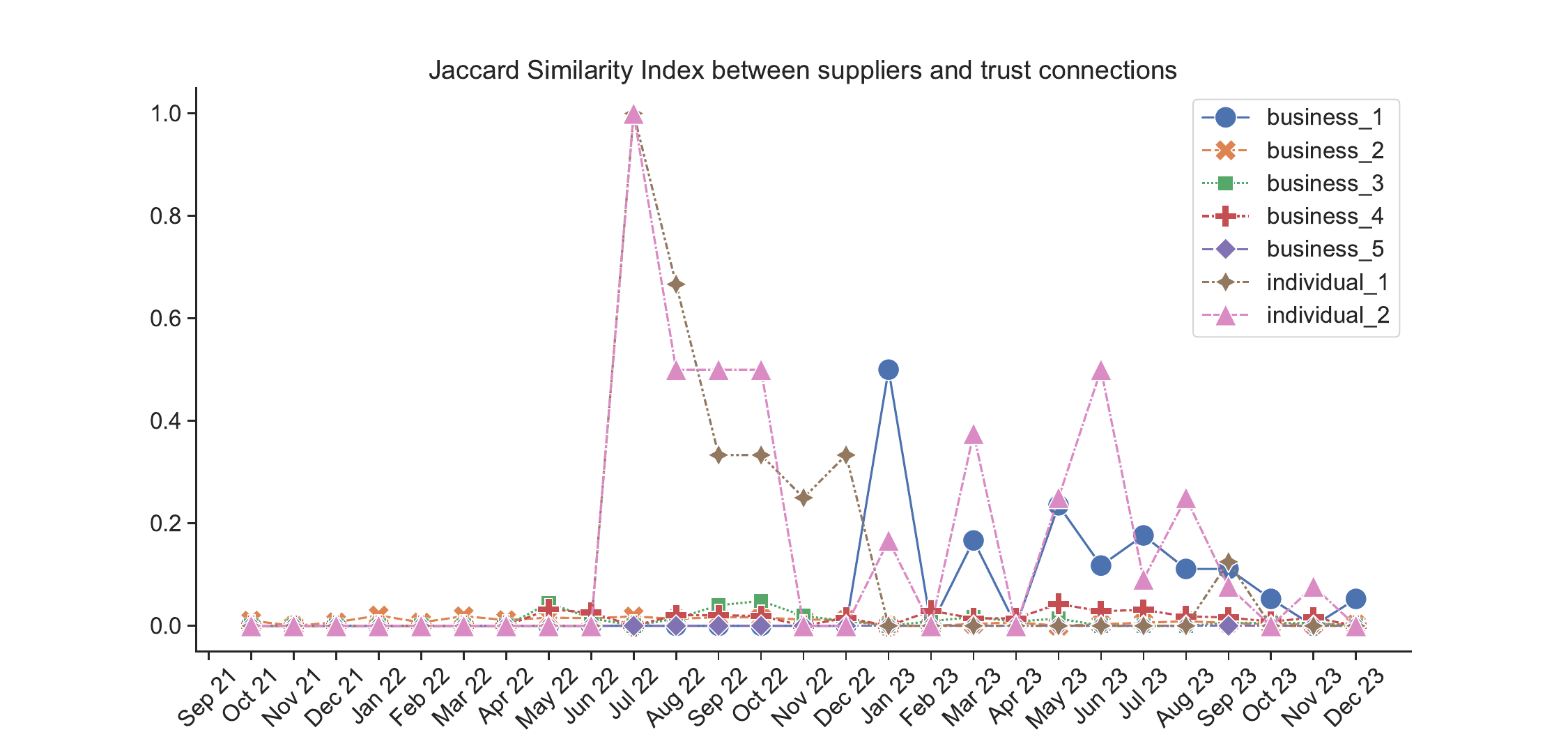}
  \caption{Jaccard Similarity Index between suppliers and trust connections of selected profiles. In the case of non-business individuals, \textit{suppliers} are those to whom such individuals send Circles.}
  
  \label{fig:jaccard_suppliers}
\end{figure}

In Figure \ref{fig:jaccard_clients}, it is possible to observe that JSI is higher for businesses. This may indicate that businesses used to create trust connections mostly with their clients. In Figure \ref{fig:jaccard_suppliers}, it is possible to observe that JSI is higher for non-businesses. This may indicate that non-business participants used to create trust connections mostly with businesses offering goods and services in Circles. 

Concluding, this reflects the heterogeneity in perception observed in the survey. For some of the businesses, Circles was perceived also as a "marketing tool". While for some of the participants, Circles was mostly perceived as a "means of payment" and "means of exchange".

\subsection{Social Composition}\setcurrentname{Social Composition}\label{sec:results:composition}

Another relevant insight that emerged from the survey section is the social composition of the sample. The first question of this section asked what was the interviewees’ main economic activity in terms of the amount of time in the past six months. This approach reflects the fundamental idea that time is the ultimate, fundamental scarce resource and it will ultimately determine the path of human activities~\citep{juster1990}. In our questionnaire, eighteen answers indicated wage labor as their main activity in terms of time, two participants answered with a mix of self-employment activities and housework, three interviewees are unemployed and currently preparing for work, and one is a student. 

The following question asked what are the main current or past occupations of the questionnaire’s participants. We found a diverse array of working specializations: four entrepreneurs/founders, four in marketing and communication, three professionals in the health sector, two IT specialists, one administrative and financial professional, one designer, and one bike mechanic and seller. We also have four interviewees who have different professions such as "writer and facilitator", "freelance artist and social worker", "fermentation expert, herbalist and educator", and “artist, researcher, cook, activist, producer, facilitator, organizer”. With the exception of the one bike mechanic, all the participants are professionals in the tertiary sector.

Among business owners, seven do not have waged employees, including two who employ freelancers and one who employs interns, two companies count two employees each, two a dozen, one has four employees and one has one. These rank all the interviewed companies as micro and small companies, according to the indication of the European Commission~\citep{eu_sme_definition}.

\begin{figure}[H]

  \centering
  \includegraphics[scale=0.45, center]{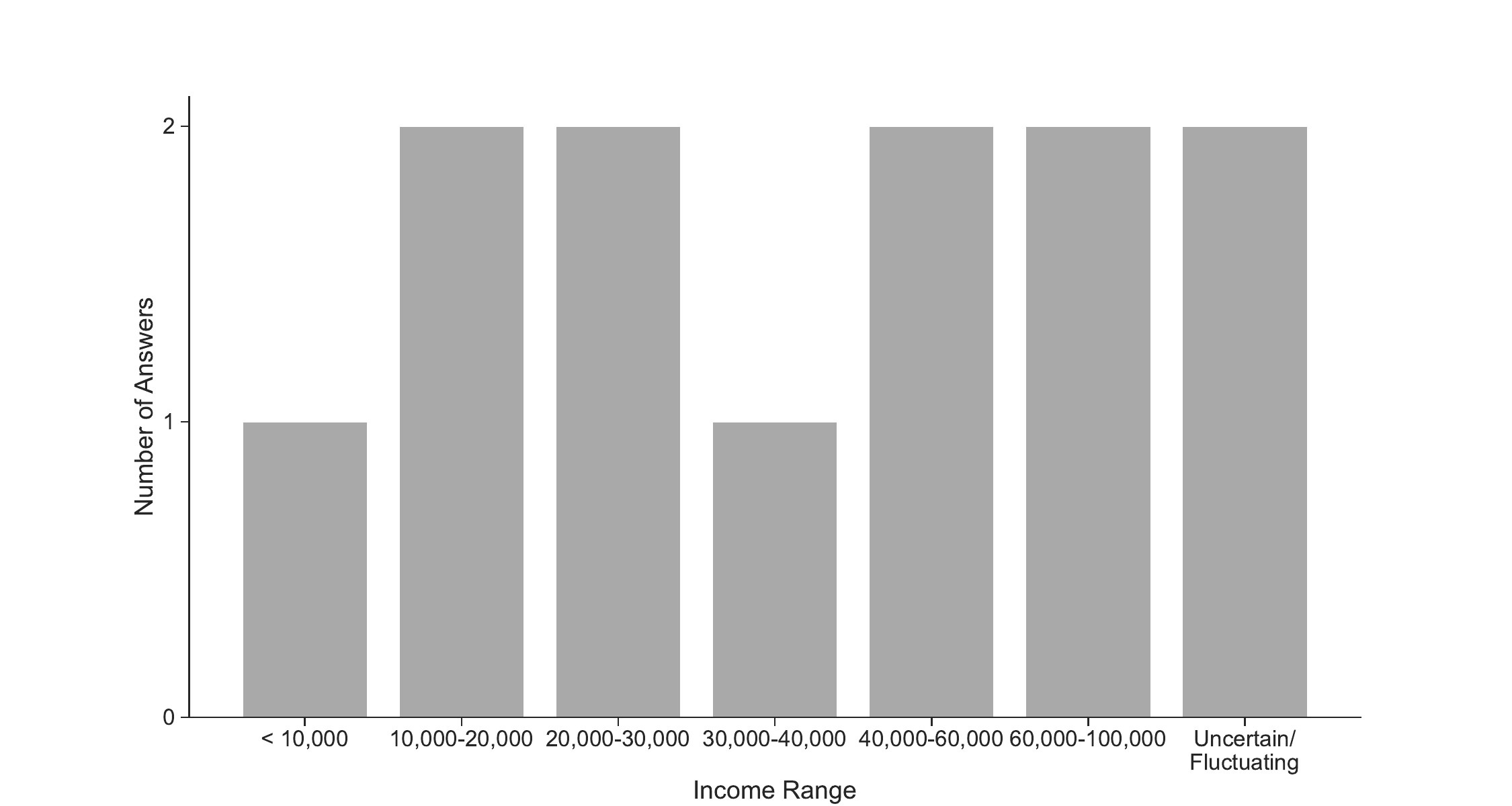}
  \caption{Income distribution among the interviewees.}
  
  \label{fig:income}
\end{figure}

The final question of this section asked the participants their gross income per year. The answers are shown in Figure \ref{fig:income}. Despite the limited sample, these data give us a glimpse into the income diversity of Circles' network. In Germany, the yearly poverty threshold in 2022 amounted to €15,000 for a “single person household”~\citep{destatis2023}, and the median annual income in 2023, in the Land of Berlin, is €43.179~\citep{stepstone2023}. Thus, our sample has one individual below the national poverty threshold and two individuals floating around it. However, six participants fall below the median annual income. Yet, no strong correlation between income and Circles participation is observed. Nonetheless, the majority of the participants expressed precarious working conditions and unstable income flows. This often underestimated aspect comes to the forefront through the granular methodology of financial diaries. In the aforementioned study led by Collins, for instance, one prevalent theme consistently identified among participants was the irregular and unpredictable nature of their income. This underscores the critical importance of delving into financial intricacies at a granular level, revealing the challenges individuals grapple with in managing their finances due to the inherent variability in their income streams.

\subsection{Challenges}\setcurrentname{Challenges}\label{sec:results:challenges}

During the process of interviews, we encountered a series of challenges. Firstly, participants’ availability and participation posed a first challenge to complete the greater number of interviews in the available time. Secondly, an additional challenge revolved around the imperative task of establishing trust within the limited time frame of the interview. This challenge stems from the inherent sensitivity of discussing one's relationship with money and the delicate aspect of financial habits and expenditure patterns. Another challenge is simply the possibility of the participants to recall wrongly the terms of their participation within Circles, especially when talking about transactions and expenses. Finally, as already mentioned, the unforeseen stop of the operations of the Circles Coop maintaining the project’s infrastructure and subsidizing the affiliated enterprises, influenced the interviews’ outcome. 

% --------------------
\section{Profiles}\setcurrentname{Profiles}\label{sec:profiles}
% --------------------

In the course of our qualitative analysis, we identified different categories of participants in Circles, both individuals and businesses. These stories can help illuminate the multifaceted impact that the project had.  

\subsection{Unemployment}\setcurrentname{Unemployment}\label{sec:profiles:unemployment}

Maria (Interview \#24, name is fictional) is a 28-year-old Italian woman coming from a middle-class family living in the North-East of Italy area. She has been living in Berlin intermittently since 2017, juggling work and pursuing her Master's degree. Her introduction to Circles came in the spring of 2022 through a friend's recommendation, and she was immediately drawn to the project's concept—a UBI issued through an innovative technology she had heard about. Being well-educated and politically engaged, Maria quickly grasped the potential of Circles and became an active member of the Berlin community, frequently participating in the monthly market and sharing her ideas in the subsequent Assembly.

During her first year in Circles, Maria worked part-time as a Marketing Manager in a German fitness company, earning a middle-class salary of 1500€ per month, falling in the lower segment of the middle class in the country~\citep{oecd2021}. However, she felt her purchasing power diminishing compared to her life in Italy, especially as the Russian-Ukrainian War triggered inflation in Germany. Maria found that participating in Circles helped her adjust her needs to the rising prices, enabling access to products that were too expensive otherwise. This gradual shift in consumption habits led her away from big supermarkets toward a local and more sustainable economic circuit.

A significant shift occurred in Maria's Circles experience when she resigned from her corporate job in May 2023 and started receiving Bürgergeld (formerly known as Arbeitslosengeld II), Germany's unemployment payment for adult job seekers (€502 per month plus rent expenses). From this point, Maria began focusing her expenses in CRC primarily on groceries, buying handmade products less frequently than before. During this period of greater need, Maria felt that the Circles UBI program played a crucial role in maintaining her quality of life. We quote here a relevant abstract from her interview: \emph{"I feel like being part of something that is taking care of me. Someone that's not my parents or myself, but a third actor that's helping and taking care. Ensure me that also in difficult times I had something good to eat and I could afford quality products."}.

Simultaneously, her unemployment condition provided more free time, prompting Maria to give back to the Circles community. She began assisting a local shop involved in Circles, MIRA Upcycling Design, with communication and social media management tasks—an ongoing collaboration at the time of writing (December 2023). Initially, Maria and the business agreed to compensate for this work in CRC. Both sides were interested in making this arrangement happen in Circles' alternative economic circuit. However, a mixture of bureaucratic problems and the unexpected closure of Circles stalled the deal, and Maria's work remained voluntary. Lastly, Maria highlighted how her experience in Circles sparked a greater interest in UBI programs, showcasing that they do not incentivize laziness but rather active engagement.

\begin{figure}[H]

  \centering
  \includegraphics[scale=0.45, center]{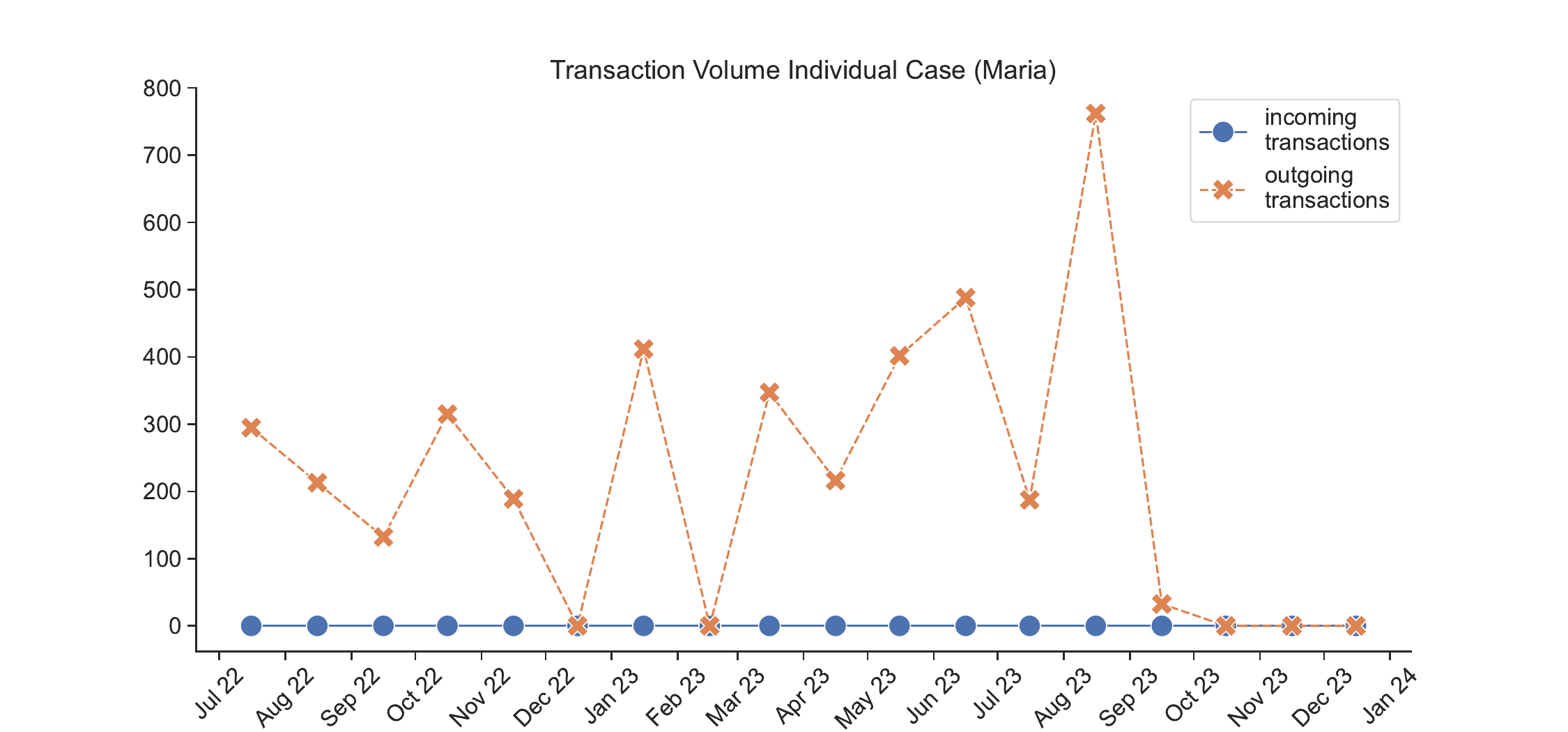}
  \caption{Total amount of outgoing transactions made by Individual Case (Maria) in Circles units per month.}
  
  \label{fig:user2_unemployed}
\end{figure}

The plot of the participant's transactions (see Figure \ref{fig:user2_unemployed}) corroborates the significant impact that Circles had during Maria's period of unemployment, commencing in May '23 and reaching its peak in September '23. Compared to the previous period, Maria started to rely more on Circles UBI system and her outgoing transactions increased. 

There are three main takeaways from this case. Firstly, it serves as evidence of how Circles UBI effectively supported an unemployed person in maintaining a good quality of life during economically unstable times. This is demonstrated by the shift in expenses after Maria became unemployed, with a focus on essential needs, particularly groceries. Secondly, in Maria's case, Circles successfully transformed the standard relationship with major food distributors in a big city, encouraging a shift toward a local and sustainable economy. Finally, Maria's active engagement in the Circles community underscores the fundamental, community-centered nature of Circles; the subject got involved in the community as soon as her life conditions changed, giving Circles more relevance in her life and providing her the necessary free time to participate.

\subsection{Casual Participants}\setcurrentname{Casual Participants}\label{sec:profiles:casual}

Among the interviewed participants, we identified an emerging category of Circles’ participants: what we may call  “casual participants”. These three profiles - interviews \#8, \#20, and \#21 - are characterized by a steady expenditure of CRC but little or no involvement in the community. Namely, two of them never participated in the Assembly and one did not join any events. Moreover, none of these profiles ever thought to contribute actively to the community. Yet, they expressed appreciation and gratitude for their participation in the program and noted how Circles unlocked access to a new category of products for them. Also, they all expressed interest and appreciation for the community aspect of Circles. Therefore, these participants are aware of the important role the community plays in the Circles network. Yet, they behave as pure "recipients" of an income program without concern for the fate of the community that allowed this program to exist.  

This fact can be read as evidence of the difficulty of changing the imagery related to this type of economic initiative. When we decouple UBI from the state, certain assumptions regarding participation in an economic system are challenged. It was not easy for all Circles participants to grasp how a contribution to the community was a contribution to UBI's system. This underscores the extent to which the conceptualization of UBI is closely intertwined with the notion of the State.

The difficulty for some Circles participants to fully comprehend the connection between their contribution to the community and the functioning of the UBI system is indicative of the entrenched imagery surrounding such socio-economic assistance models. The existence of a subsidy program for active stores inadvertently downplayed the perceived importance of individual contributions, fostering a mindset that leaned towards the notion that as long as the subsidy program and the cooperative persisted, Circles UBI would endure. From this perspective, the Circles Coop and its monetary decisions interpret a similar role to what the State usually does in such contexts. 

Moreover, the temporal dimension played a significant role. Understanding the nuances of a Circles system required time, and such participants with a lower engagement level joined the platform between six months and a year before the fall of 2023. The unexpected cessation of the project in Berlin further curtailed opportunities for participants to become actively involved in the community. This temporal aspect, combined with the abrupt shutdown, added layers of uncertainties to their engagement and obstacles to the comprehension of the Circles UBI experiment. Various factors contributed to the existence of this category of participants within Circles. It is crucial to acknowledge that, as the Berlin local pilot evolved, these "casual participants" played a significant role.

A concluding observation: this category underscores how, in a community-based experiment like the one being examined, the unconditionality feature of UBI must be interpreted in a slightly nuanced manner. More precisely, unconditionality shouldn't be linked to a lack of participation, a distinction that might exist in a State-based version of a UBI experiment.

\subsection{Business Case 1}\setcurrentname{Business Case 1}\label{sec:results:profiles:business1}
%The case of MIRA Design}

M. is a 39-year-old Brazilian-Italian woman living in Berlin. She has a social business named MIRA Design which designs and produces upcycled fashion items. Founded in 2006 in Curitiba, Brazil, in 2013 MIRA expanded into Berlin with the line ‘Proudly made in Bangladesh’. MIRA made in Bangladesh outsources textile clipping waste from Bangladesh’s ready-made garment industry (RMG) and brings them into the hands of the skilled craftswomen in rural Rangpur, Northern Bangladesh. Then, MIRA sells products in Berlin and worldwide through the website. 

MIRA's relationship with Circles began through another local Berlin shop, Eddie's Shop. Their story is an example of how a synergy between businesses has been partially translated from the real economy to the  Circles currency system. In the early months of 2022,  Eddie's Shop, a member of the network since 2021, started selling MIRA's products in the store, accepting commissions for them. Eddie's was purchasing MIRA’s products in EUR and selling them both in EUR and CRC with success. For this reason, in the fall of ‘22, MIRA decided to actively join the Circles community and sell its products directly in the marketplace. Around the same months, Eddie's Shop exited the subsidy program and withdrew from the Circles network due to the unsustainability of the business, preventing this synergy from surviving within Circles. The idea of establishing network economic linkages between Circles members has been a key strategic point of the Circles cooperative. While the linkage between Eddie's and MIRA stopped prematurely, we will examine a more successful linkage below.

\begin{figure}[H]

  \centering
  \includegraphics[scale=0.45, center]{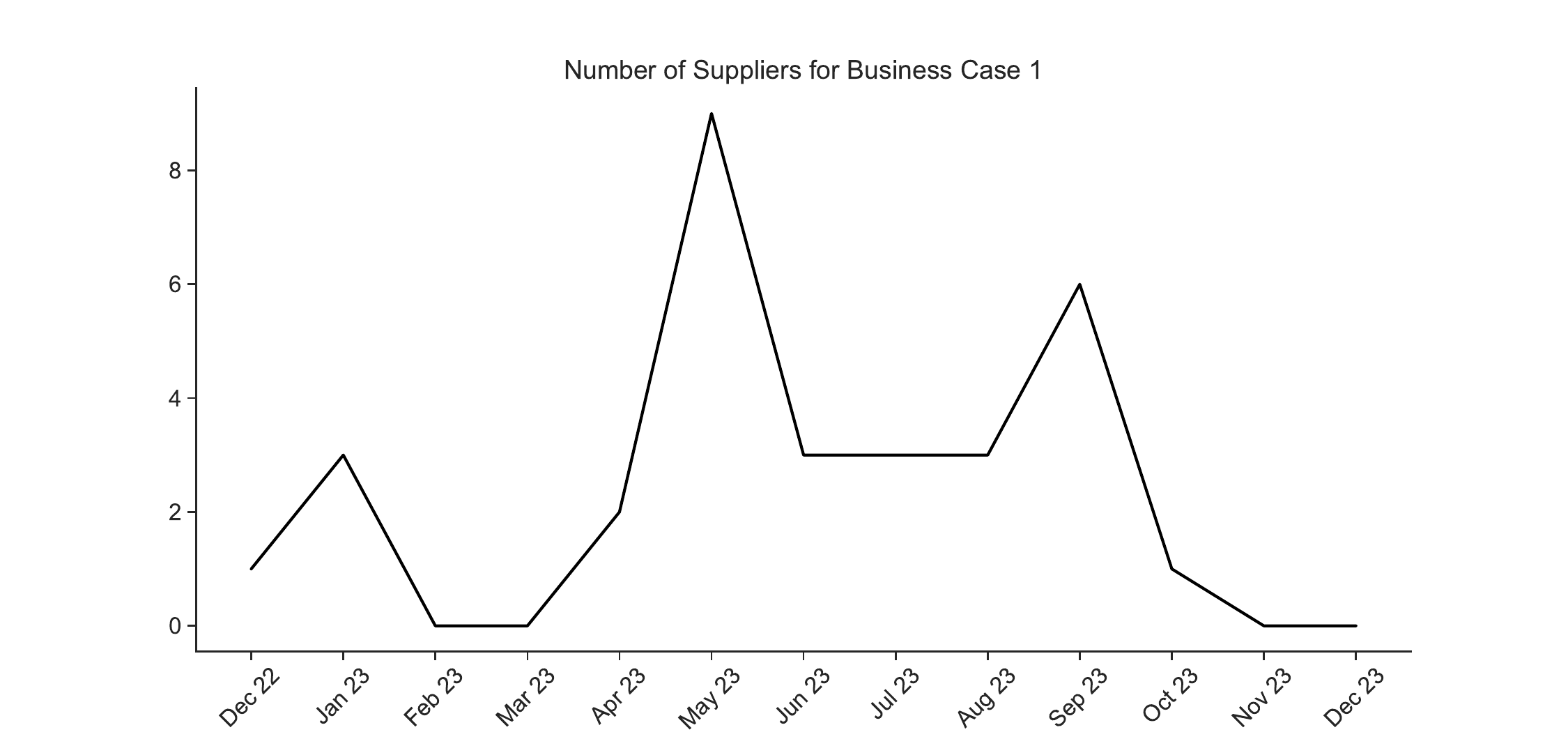}
  \caption{Number of suppliers of Business 1 in Circles network per month.}
  
  \label{fig:business1_transactions}
\end{figure}

MIRA's story in Circles is also insightful for another reason: it was an active part of the Circles network for three months before subsequently joining the subsidy program, unlike most of the businesses involved, who have been in the subsidy program since the beginning of their Circles experience. In the short term, the business sustained itself without resorting to the option of exchanging the CRC token for EUR, through the subsidy. The impact of the subsidy can be traced to the spike of incoming transactions around April ‘23 in Figure \ref{fig:business1_transactions}, when the subsidy program started. Also, the subsequent peak in outgoing transactions can be explained by the fact that MIRA’s owner started to exchange their CRC in EUR. 

\begin{figure}[H]

  \centering
  \includegraphics[scale=0.45, center]{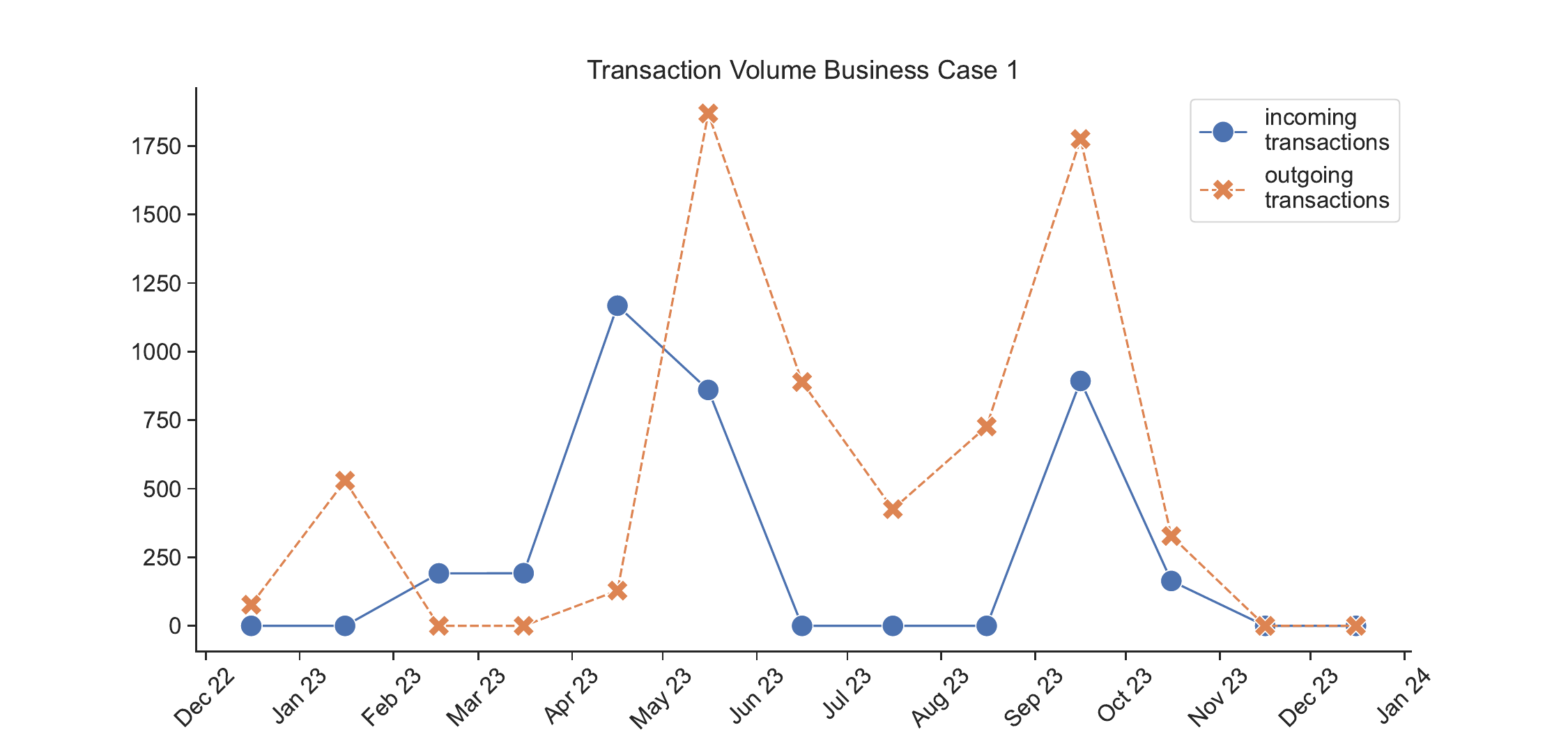}
  \caption{Total amount of transactions of Business 1 in Circles units per month.}
  
  \label{fig:business1_suppliers}
\end{figure}

Finally, MIRA also revealed the hurdles that it encountered during its period in the network while attempting to integrate Circles into its business model. A key issue revealed in the interview was the limited availability of desired services for fostering MIRA’s business activities within the Circles marketplace, hindering the business's ability to effectively utilize the currency. MIRA was looking for marketing, graphic design, and video editing skills on the market and it struggled to find them. This mismatch resulted in the accumulation of CRC without sufficient opportunities for expenditure (see Figure \ref{fig:business1_suppliers}. ) (with the notable exception of Maria’s case presented above). Moreover, MIRA’s owner found little or no time to explore the network’s offer, even more so considering the abrupt project’s shutdown. 

\subsection{Business Case 2}\setcurrentname{Business Case 2}\label{sec:results:profiles:business2}

Local Community Currency Systems can provide a fertile ground for the flourishing of small and medium enterprises~\citep{cepel2019local}. This positive impact has been evident in experiments conducted in various locations, such as Argentina~\citep{colacelli2005secondary} and Canada~\citep{wheatley2011calgary}. As revealed in the interview, the narrative of QueerQuaff serves as a compelling example within Circles, illustrating the tangible benefits of such systems for the growth of local businesses.

QueerQuaff is a feminist and queer beer project based in Berlin, which activities got bootstrapped thanks to their participation in Circles. It presents itself as “A space for learning,experimentation, solidarity and rediscovery of feminist beer history”. The two project’s founders first met in 2020 for an artistic residency in the Spreewald, a cultural hub in Berlin growing out of the ruins of an East Germany abandoned amusement park. The two found some wild hosp foraged there and decided to embark on the process of fermentation. The idea gained traction, and the two started producing a bigger batch in a house project (\textit{Hausproject}) in Berlin known as H48. Around the spring of 2021, QueerQuaff’s team met with members of the Circles cooperative and they recognized in each other similar values. The synergy resulted in an initial investment from Circles in QueerQuaff’s new production, which allowed them to obtain their necessary means of production, such as a professional kitchen. This investment is among the first in the history of the pilot in Berlin and it precedes the establishment of the subsidy program. QueerQuaff organized a crowdfunding accepting CRC tokens that then were exchanged through the cooperative in Euro. In this way, the bootstrap of a local business helped the Circles network in gaining momentum. 

\subsection{Economic Synergy}\setcurrentname{Economic Synergy}\label{sec:results:profiles:synergy}

As stated previously, a key strategy to foster and expand the Circles network concerned the creation of economic linkages between businesses in the network. Economic linkages denote the intricate and reciprocal relationships among diverse sectors, industries, or economic entities within a given economic framework~\citep{hirschman}. These connections encompass both forward linkages, delineating the directional flow of goods or services from initial production stages towards ultimate consumption, and backward linkages, which elucidate the dependencies on inputs or raw materials provided by antecedent sectors~\citep{tenraa2020}. Through our interviews, we noticed the emergence of several fruitful linkages through subsided partners in the network. Specifically, within the context of Circles, we focus on Business-to-Business interactions. These relationships have been proven to be effective in expanding production capabilities and increasing sales through business collaborations. 

A significant linkage has been established between RealRadix and FarmerConnection, two Berlin-based food and beverage businesses with a focus on sustainability. RealRadix is an authentic, handcrafted, and zero-waste food company dedicated to reconnecting people with food. On the other hand, FarmerConnection is a local and sustainable enterprise that distributes fruits and vegetables from the neighboring Brandenburg region to Berlin.

Among FarmerConnection's array of products, they deal with green tomatoes, which, due to their acidity and other chemical characteristics, cannot be consumed directly. However, recognizing their potential value, FarmerConnection sought an innovative solution to avoid wastage and enlisted RealRadix for the production of green tomato jam. The founder of FarmerConnection, already a significant consumer of RealRadix's products, decided to entrust them with the task.  The green tomato jam, a result of this collaboration, has been successfully finalized and is now available for sale and distribution through FarmerConnection's website and services. This strategic partnership, mediated by CRC-based transactions not only addresses the challenge of food waste but also enhances the product offerings for both businesses. It has to be said, however, that FarmerConnection, as stated by its founder, has already been involved in those kinds of strategic partnerships before its participation in the Circles network. 

Another synergistic collaboration within the Circles network involved a fermentation business Food Magic and the ceramic shop Tinted Life. As revealed in the interview with the co-founder of Food Magic, Tinted Life crafted a personalized set of pottery plates and cutlery designed specifically for use during Food Magic's culinary experiences and fermentation workshops. These unique artifacts were utilized to enhance the overall dining experience provided by Food Magic's services. In return for these customized pieces, Food Magic paid Tinted Life using CRC (Circles currency). Subsequently, Tinted Life used these earned CRC on the Circles marketplace. The co-founder of Food Magic emphasized that this partnership, although driven by a strategic vision, was also influenced by the abundance of CRC at her disposal and the motivation to engage with the Circles community. 

The creation of economic linkages serves to fortify the autonomy of the Circles economic network by diminishing the need for exchanging CRC into Euros, thereby fostering sustained mutual growth among businesses operating within Circles. Additionally, it alleviates the dependence of businesses on the Cooperative's subsidy program.

% -----------------------------------------
\section{Discussion}\setcurrentname{Discussion}\label{sec:discussion}
% -----------------------------------------

In this paragraph, we report the main critical issues discussed in the last assemblies of Circles Coop and the related proposed solutions. The experience of Circles Coop and its Berlin pilot shows that any autonomous group willing to implement basic income from the bottom up needs to be economically self-reliant. In this way, it can continue to reproduce itself as a body that coordinates and expands the pilot. 

The experimental nature of Circles meant a high risk for the businesses involved. The subsidy and redemption system was supposed to offset this risk. Nevertheless, this strategy was limited because Circles Cooop had no secure revenue. One of the outcomes was an unbalance between givers and receivers in the Circles network. In other words, the number of suppliers and their involvement was insufficient for the creation of remarkable and long-lasting economic synergies.

The first proposed solution by the local activator groups (Community Activation Teams, CATs) was to start implementing credit operations. In this way, Circles Coop could offer loans to the Circles business network at zero or low-interest rates. This financial service could eventually target specific businesses by trying to solve the issues of supply-chain linkages more actively. Furthermore, the repayment of loans could give the cooperative enough revenue to cover its costs. 

A second proposed solution was a profit-sharing model inspired by the principles of \textit{Islamic finance}. In this framework, the cooperative would invest in local businesses as a shareholder. This approach integrated with a subsidy program was inspired by economic local protectionism which worked to develop industrial capabilities in various countries ~\citep{chang2002}. This would have been intentionally in contrast with the more recent ordoliberal ideology which argues for deregulation of markets and privatisation of public and common goods. 

According to the organizers, both discussed solutions could have allowed the cooperative to issue and manage a local currency in the form of a basic income. The adoption of a local currency was mostly inspired by the successful cases of Banco Palmas, Niteroi, and Maricá, in Brazil ~\citep{basicincomebrasil, cernev2016}. Albeit these two interventions enjoy direct local government support, for the organizers the prospect of autonomous credit cooperatives as the vehicle for introducing basic income remained an interesting horizon.

% -----------------------------------------
\section{Conclusion}\setcurrentname{Conclusion}\label{sec:conclusion}
% -----------------------------------------

Our qualitative investigation of the impact of the Berlin pilot on Circles participants has shed light on crucial facets of this unique blockchain-based framework. By employing a financial-diaries-inspired methodology, we delved into the experiences of twenty-five participants, complementing previous quantitative analyses with narratives and nuanced insights.

Our findings underscore the pivotal role of community as the driving force behind the Circles project. Participants' narratives consistently highlighted the profound influence of interpersonal dynamics, mutual support, and shared values, revealing that the experience of Circles is deeply intertwined with the strength of its community.

Moreover, we addressed the challenges posed by technical instabilities and intricacies within the Circles system. While acknowledging these hurdles, participants showcased resilience and adaptability, demonstrating how the community effectively navigated and mitigated these challenges. This resilience not only emphasized the project's robustness but also provided valuable insights for future iterations and similar endeavors.

An essential aspect illuminated through our study is the diverse levels of participation and engagement within Circles. The decentralized and open nature of the project allows for varying degrees of involvement, accommodating participants with different preferences and comfort levels. This flexibility contributes to the project's inclusivity, attracting individuals with diverse needs and preferences. 

The identification of five distinct profiles among participants illustrates precisely this dynamic, namely the multifaceted impacts of Circles on individuals' lives. Each profile delineates a unique narrative, elucidating how Circles UBI helped or sustained different categories of individuals and businesses in different methods. Particularly, the successful stories of economic linkages suggest a path for Local Community Currencies that can be repurposed in future iterations. 

In a broader sense, our study bridges academic inquiry with a pragmatic and strategic outlook. By offering a comprehensive understanding of Circles' impact, we aspire to provide valuable insights for Circles and similar pioneering projects as they navigate their future trajectories. Our findings underscore the intricate interplay between elements such as community dynamics, technical challenges, diverse levels of participation, and the tangible, varied impacts experienced by individuals. In future experiments of this kind, all of these components will need to be considered. 

% %%%%%%%%%%%%%%%%%%%%%%%%%%%%%%%%%%%%%%%%%%%%%%%%%%%%%%%%%%
% %%%%%%%%%%%%%%%%%%%%%%%%%%%%%%%%%%%%%%%%%%%%%%%%%%%%%%%%%%
% REFERENCES SECTION
% %%%%%%%%%%%%%%%%%%%%%%%%%%%%%%%%%%%%%%%%%%%%%%%%%%%%%%%%%%
% %%%%%%%%%%%%%%%%%%%%%%%%%%%%%%%%%%%%%%%%%%%%%%%%%%%%%%%%%%
\medskip

% \bibliography{bibliography} 

\newpage

% %%%%%%%%%%%%%%%%%%%%%%%%%%%%%%%%%%%%%%%%%%%%%%%%%%%%%%%%%%
% %%%%%%%%%%%%%%%%%%%%%%%%%%%%%%%%%%%%%%%%%%%%%%%%%%%%%%%%%%
\section*{Figures}\setcurrentname{Figures}\label{sec:figures}
% %%%%%%%%%%%%%%%%%%%%%%%%%%%%%%%%%%%%%%%%%%%%%%%%%%%%%%%%%%
% %%%%%%%%%%%%%%%%%%%%%%%%%%%%%%%%%%%%%%%%%%%%%%%%%%%%%%%%%%

% %%%%%%%%%%%%%%%%%%%%%%%%%%%%%%%%%%%%%%%%%%%%%%%%%%%%%%%%%%
% %%%%%%%%%%%%%%%%%%%%%%%%%%%%%%%%%%%%%%%%%%%%%%%%%%%%%%%%%%
\section*{Appendix}\setcurrentname{Appendix}\label{sec:appendix}
% %%%%%%%%%%%%%%%%%%%%%%%%%%%%%%%%%%%%%%%%%%%%%%%%%%%%%%%%%%
% %%%%%%%%%%%%%%%%%%%%%%%%%%%%%%%%%%%%%%%%%%%%%%%%%%%%%%%%%%

\begin{enumerate}    
  \item \textbf{User details}
  \begin{enumerate}
    \item Name, Surname, Circles Surname
    \item Individual Wallet Address, Business Wallet Address, Email Address
    \item Phone Numbe, Age, Sex
    \item Gender, Nationality
    \item \textbf{Specific Information}
    \begin{enumerate}
    \item Within Circles network, are you an individual or a business?
    \end{enumerate}
    \item \textbf{Business details}
    \begin{enumerate}
    \item What is your Business Name? 
    \item What is your Business Address?
    \item What is your Business Sector Code?
    \item How long have you been part of Circles as business?
    \item How often do you use Circles as business?
    \item On average, how much do you spend in a month as business?
    \end{enumerate}
    \end{enumerate}

  \item \textbf{Economic Network}
  \begin{enumerate}
    \item How long have you been part of Circles as individual?
    \item How often do you use Circles as individual?
    \item On average, how much do you spend in a month as individual?
    \item Did your frequency of usage change across the time you have been a Circles user? If so, why?
    \item What do you use Circles for? What kind of goods you buy? (Be as broad and specific as you can)
    \item Would you say you use Circles to buy products and services that you would otherwise not buy? If so, in what?
    \item Would you say using Circles changed your habits of consumption?
    \item Do you share your Circles account with others?
    \item How many people do you know in Circles?
    \item Where do you spend Circles?
    \item Do you use Circles more with businesses or with private people?
    \item Have you ever offered your own products and services within the Circles community? If yes / not, why?
    \item Have you encountered instances where Circles users appeared to have buy products with CRC to resell in EUR?
    \item Have you received any feedback from customers or other businesses regarding misuse of Circles within your ecosystem?
    \end{enumerate}

  \item \textbf{Knowledge \& Perception of Circles}
  \begin{enumerate}
    \item How did you know about Circles? 
    \item What motivated you to join the Circles Network?
    \item What part of Circles are you most interested in?
    \item \textbf{Use of Circles}
    \begin{enumerate}
    \item How do you perceive Circles?
    \end{enumerate}
    \item How often do you use Circles as business?
    \item Did you take part in Circles’ events?
     \item \textbf{Circles Events Participation}
    \begin{enumerate}
    \item Which ones did you go? 
    \item How often did you go?
    \item What do you do when joining the Circles event?
    \end{enumerate}
    \item \textbf{Future of Circles}
  \begin{enumerate}
    \item Has your vision/opinion on Circles changed since you’re part of the program? How? 
    \item What does it mean to you to be a recipient of a basic income?
    \end{enumerate}
    \end{enumerate}
    
  \item \textbf{Socio-economic position}
  \begin{enumerate}
    \item Over the past six months, what was your main economic activity in terms of the amount of time you spent on it? 
    \item What is your past and/or current occupation?
    \item How many employees does your company have? (for owners and self-employed)
    \item What is your gross income per year in Euro?
    \item Do you or your family own property?
    \end{enumerate}

  \item \textbf{Satisfaction}
  \begin{enumerate}
    \item How much do you feel satisfied about the participation to Circles UBI project? 
    \item How much your general well-being got improved thanks to the participation to CIrcles UBI project?
    \item How likely would you recommend the participation to Circles UBI to a friend? 
    \end{enumerate}
    
\end{enumerate}

% ==========================
% ==========================
% ==========================

\end{document}